\documentclass[10pt, letterpaper, twocolumn]{article}

\usepackage{ol2}
\usepackage[draft]{hyperref}
\usepackage{amsmath}
\usepackage{graphicx}
\usepackage{epstopdf}
\begin{document}

\twocolumn[

\title{
Single-shot spectra of temporally selected micropulses from a mid-infrared
free-electron laser by upconversion
}

\author{Xiaolong Wang, Takashi Nakajima$^1$, Heishun Zen, Toshiteru Kii, 
and Hideaki Ohgaki$^2$}

\address{
Institute of Advanced Energy, Kyoto University, 
Gokasho, Uji, Kyoto 611-0011, Japan\\
$^1$e-mail: t-nakajima@iae.kyoto-u.ac.jp\\ 
$^2$e-mail: ohgaki@iae.kyoto-u.ac.jp 
}

\begin{abstract}
We demonstrate the measurement of single-shot spectra of temporally selected 
micropulses from a mid-infrared (MIR) free-electron laser (FEL) 
by upconversion. 
We achieve the upconversion of FEL pulses at 11 $\mu$m using 
externally synchronized Nd:YAG or microchip laser pulses at 1064 nm 
to produce sum-frequency mixing (SFM) signals at 970 nm, which 
are detected by a compact CCD spectrometer without an intensifier. 
Our experimental system is very cost-effective, and allows us 
to obtain the laser spectra of selected micropulses 
at any temporal position within a single macropulse from an oscillator-type 
FEL. 
\end{abstract}



]


Since the first realization of lasing in 1976 \cite{Deacon:1977zz}, FELs 
have been attracting lots of interests in various research areas due to 
their large tunabilities in wavelength \cite{OShea:2001gv}. 
In order to apply the FEL pulses for processes involving resonances 
\cite{Zimdars:1993tf}, 
precise knowledge of laser spectrum as well as pulse duration is 
particularly important. Moreover, since the lasing mechanism of FELs is
essentially different from that of conventional lasers, such information
on FEL pulses themselves sheds crucial light for the profound understanding 
of lasing mechanism of FELs. This can be particularly true for the 
oscillator-type FELs which exhibit a dual pulse structure: 
A macropulse with a duration of $\mu$s to ms contains thousands of 
micropulses with a duration of fs to ps at a time interval of hundreds of 
ps to several ns. We point out that most of the FELs working at the
MIR wavelength range are the oscillator-type. 
For this reason, the pulse measurements of MIR FELs 
are almost always carried out by averaging over many macropulses, 
each of which contains thousands of micropulses. 
For instance, even a single-macropulse measurement of the FEL spectra, 
which is usually considered to be easy, is not so easy in the MIR range, 
and it requires the use of an array-type MCT (mercury cadmium telluride) or
pyroelectric photodetectors whose resolution is limited by the low pixel 
density (typically 64-256 elements). 
Moreover, although the use of such array-type photodetectors allows us 
to obtain the MIR FEL spectrum for a single macropulse, it is still an average 
over thousands of micropulses within a single macropulse, since the 
time response of those detectors are rather slow. 
As a result, although the MIR FEL spectra of temporally selected 
micropulse(s) are much more valuable to monitor the laser spectra during
the spectroscopic experiments and to understand the detail of the lasing 
process, such study does not exist in the literature. 

The purpose of this paper is to report the measurement of single-shot spectra, 
for the first time to our knowledge, of 
{\it temporally selected micropulse(s)} from an MIR FEL by upconversion 
using an externally synchronized laser source at 1064 nm. 
Although the upconversion of MIR pulses with visible or near-infrared 
(NIR) pulses was first reported quite some time ago 
\cite{Shah:1988wl,Heilweil:1989tw}, it is much more recent 
that the idea has been applied for the MIR 
\cite{Kubarych:2005io, DeCamp:2005va, Baiz:2011tf, Zhu:2012wq}.
Related to the present work 
we note that the upconversion of THz radiation from an FEL by a NIR 
continuous-wave laser has been recently reported \cite{Wijnen:2010tb}.

We would like to note that whether the upconversion technique works 
equally well for the case of MIR FELs is not {\it a priori} 
obvious due to the following reasons: 
First, our MIR pulses from the FEL form a train of several thousands of 
micropulses, while in all the works mentioned above the MIR pulses are 
isolated. 
The peak intensity one can safely use for an isolated pulse may be too much 
for a pulse train, and can damage a nonlinear crystal for upconversion. 
Second, we do the synchronization of two independent lasers and hence 
there is some timing jitters, while in all the above works 
both MIR and NIR pulses are from the single light source 
and hence the synchronization is nearly perfect. 
To reduce the problem of timing jitters we employ the ns or sub-ns lasers 
at the expense of wasting most of the pulse energy due to the mismatch 
of the pulse durations between the MIR and NIR pulses. 
Third, due to the use of a NIR laser source at 1064 nm the SFM signals 
in our case appear around 970 nm where the sensitivity of Si-based 
CCD photodetectors is usually quite low ($<$ 10 \% of its maximum value 
which typically appears around 500-700 nm), while the sensitivity of the 
detector is much higher if one uses, for example, a Ti:Sapphire laser 
at the expense of its cost. 

\begin{figure}
\centerline{
\includegraphics[width=8cm]{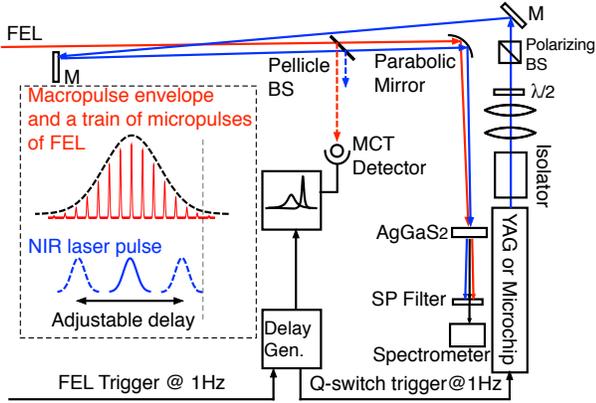}
}
\caption{
(Color online) Experimental setup. 
}
\label{fig:experiment_plan}
\end{figure}

The experimental setup is shown in Fig. \ref{fig:experiment_plan}. 
The Kyoto University free-electron laser (KU-FEL) operates at 11 $\mu$m 
with a repetition rate of 1 Hz \cite{kufel}. 
Each macropulse has a duration of $\sim$1.5 $\mu$s and contains 
several thousands of micropulses with a duration of about 0.7 ps and 
an interval of 350 ps between them. 
As for the 1064 nm laser we employ either the Q-switched Nd:YAG 
(LOTIS TII, model LS-2136, multi-mode, 20 ns pulse duration, 50 ns jitter) or 
actively Q-switched microchip lasers (Standa, STANDA-Q1, single-mode, 
0.8 ns pulse duration, $<350$ ps jitter). In either case the Q-switch of the 1064 nm laser is triggered by the 
emission timing of KU-FEL with some controlled delay through the 
delay generator (SRI, DG645). 
Note that the timing jitter of the NIR lasers is not a serious problem
for our case, since their pulse durations are larger than the interval between 
the micropulses of FEL.

Unless otherwise noted, we employ the Nd:YAG laser for upconversion. 
A half-wave plate and a polarizing beam splitter are introduced for the 
1064 nm pulse so that its polarization axis becomes parallel to that 
of the FEL pulse. 
Then, both beams are focused onto the AgGaS$_2$ crystal 
(type I, 2 mm thickness, 37 degrees cut angle) using a gold-coated 
off-axis parabolic mirror with a focal length of 15 cm. 
The SFM bandwidth for the 2 mm crystal is about 0.7 $\mu$m 
at 11 $\mu$m, which is much larger than the bandwidth of the retrieved 
FEL spectrum, $\sim$0.3 $\mu$m (see Fig. 3(d)), which ensures that 
our choice of the crystal thickness is appropriate.
The energy of FEL pulses is 4 mJ per macropulse, while that of the 
1064 nm pulses is 5 mJ per pulse at the AgGaS$_2$ crystal. 

A rough estimation shows that each macropulse of KU-FEL contains approximately 
4000 micropulses and hence the energy of each micropulse is about 1 $\mu$J. 
Since the diameters of the two beams at the crystal are measured to be 
about 1 mm for both, the peak intensities of the FEL micropulse 
and 1064 nm pulse are about 130 MW/cm$^{2}$ and 25 MW/cm$^{2}$, respectively. 
The relatively low peak intensities prevent not only damage 
on the crystal but also distortion of the spectrum by self-phase modulation 
in the crystal.
The two incident beams have a small cross angle of about 2.5$^{\circ}$ 
so that we can spatially separate the SFM signals from the incident 
beams. After the AgGaS$_2$ crystal we place a shortpass filter 
with a cut-off wavelength at 1000 nm to block the 1064 nm pulse. 
For the detection of SFM signals, we employ a commercial compact 
spectrometer (OceanOptics, model HR4000CG-UV-NIR) with a slit width of 
5 $\mu$m and grating of 300 lines/mm (0.75 nm spectral resolution). 
The spectrometer is also synchronized with the laser system. 
As a reference we pick up a small portion of the FEL pulse 
by a Pellicle beam splitter to monitor the temporal shape and energy 
of the macropulse by an MCT detector, 
and store the data with a 2 channel digital oscilloscope. 

\begin{figure}
\centerline{
\includegraphics[width=8cm]{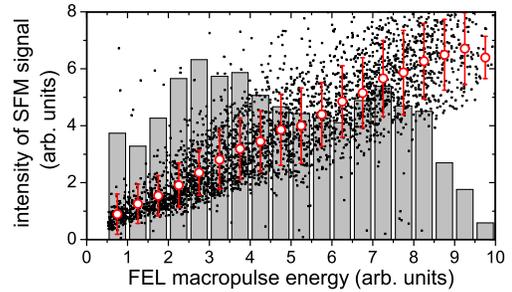}
}
\caption{
(Color online) Correlation between the intensity of SFM signals 
and FEL macropulse energy. 
\emph{Small black dots}: Raw data; 
\emph{Grey columns}: Frequency of events ; 
\emph{Red circles and error bars}: Averaged SFM signal intensity and error bars. 
}
\label{fig:intensitycorrelation}
\end{figure}

As a first test we have measured the SFM signal intensity 
as a function of FEL macropulse energy. 
The results are shown in Fig. \ref{fig:intensitycorrelation}. 
The shot-to-shot change of the 1064 nm pulse energy is $<$ 20\%.  
Regardless of the large shot-to-shot fluctuation of FEL macropulse 
energies, we can clearly see the nearly linear dependence, as we expect. 
Recall that the SFM signal intensity is proportional to the 
micropulse intensity (and with a good approximation macropulse energy) of FEL. 
By calibrating the spectrometer sensitivity at $\sim$970 nm, 
we estimate the SFM efficiency to be about 0.2\% $\sim$ 0.6\%.

\begin{figure}
\centerline{
\includegraphics[width=8cm]{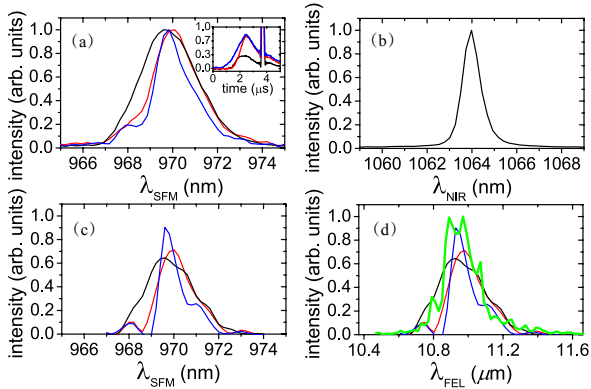}
}
\caption{
(Color online) Retrieval of the spectra of FEL micropulses by 
upconversion. 
(a) Raw spectra of SFM signals. Three examples, normalized by their peak 
values, are shown. The inset shows the corresponding macropulse shapes. 
where the narrow pulse appearing at 3.5 $\mu$s is the 1064 nm pulse. 
(b) Spectrum of the microchip laser.
(c) Deconvolved spectra of SFM signals from raw spectra in (a).
(d) Retrieved MIR spectra of the temporally selected FEL micropulses
and the MIR spectrum (thick green line) taken by the scanning-type 
monochromator (amplitude not in scale).
}
\label{fig:spectrumdeconvolution}
\end{figure}

In Fig. \ref{fig:spectrumdeconvolution}(a) we present a few raw spectra 
of the SFM signals, each of which has been obtained by upconversion of 
temporally selected micropulses of the different macropulse. 
The timing of the 1064 nm pulses is set to be at about 1.2 $\mu$s 
after the peak of the FEL macropulse envelope 
(see the inset in Fig. \ref{fig:spectrumdeconvolution}(a)). 
Because the pulse duration of the Nd:YAG laser is 20 ns and the time
interval between the FEL micropulses is 350 ps, the number of 
FEL micropulses we have selected for upconversion is about 57. 
In order to retrieve the MIR FEL micropulse spectra, we must perform 
the deconvolution. 
We have found that the spectral profiles of the multi-mode 
Nd:YAG ($\sim$0.45 nm bandwidth) and single-mode 
microchip ($< 3.7 \times 10^{-3}$ nm bandwidth) lasers measured by our 
spectrometer are almost identical, as shown in 
Fig. \ref{fig:spectrumdeconvolution}(b). 
This implies that the spectral resolution of our spectrometer, 0.75 nm, 
is much worse than the linewidths of the Nd:YAG and microchip lasers, 
and the spectral profile shown in Fig. \ref{fig:spectrumdeconvolution}(b) 
is nothing but the instrumental function of the spectrometer, which we use 
for deconvolution. 
The deconvolved SFM spectra are shown in 
Fig. \ref{fig:spectrumdeconvolution}(c). 
Once the SFM spectra have been obtained by deconvolution, 
we can retrieve the spectra of the temporally selected FEL micropulses, 
$\lambda_{\rm\scriptstyle FEL}$, by using the relation of 
$\lambda_{\rm\scriptstyle FEL} 
= ( \lambda_{\rm\scriptstyle SFM}^{-1} 
- \lambda_{\rm\scriptstyle NIR}^{-1} )^{-1}$, where
$\lambda_{\rm\scriptstyle SFM}$ and $\lambda_{\rm\scriptstyle NIR}$ 
are the wavelengths of the SFM signal and NIR pulse, respectively. 
The retrieved spectra of temporally selected FEL micropulses are shown 
in Fig. \ref{fig:spectrumdeconvolution}(d). 
For comparison we also show in Fig. \ref{fig:spectrumdeconvolution}(d) 
the FEL spectrum taken by a scanning-type monochromator, which appears noisy because of the shot-to-shot fluctuation of the FEL pulse energy.

In order to improve the temporal resolution, 
we now replace the 20 ns Nd:YAG laser by the 0.8 ns microchip laser.
Since the effective pulse energy of the microchip laser is about half 
of that of the Nd:YAG laser, we move the AgGaS$_2$ crystal closer to the 
focal point, and the diameters of the FEL and 1064 nm beams are 
reduced to 0.5 mm, which results in the peak intensities of 
500 MW/cm$^{2}$ and 50 MW/cm$^{2}$, respectively. 
Regardless of such high peak intensities we find that one hour laser 
irradiation at 1 Hz does not induce any visible damage on and in the crystal. 

\begin{figure}
\centerline{
\includegraphics[width=5.5cm]{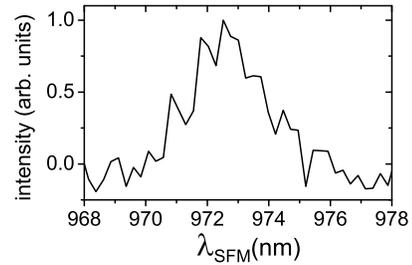}
}
\caption{
Raw spectrum of the SFM signal with the microchip laser. 
}
\label{fig:microchipspectrum}
\end{figure}

In Fig. \ref{fig:microchipspectrum} we show a raw spectrum of the SFM signal 
obtained with the microchip laser. 
Note that only $\sim$2 micropulses of FEL are selected for upconversion. 
Although the SFM signals become much weaker, we believe that 
there is a lot of room for improvement through the replacement 
of the spectrometer by the one with higher sensitivity ($> \times$20) and
the grating by the one with much better ($> \times$5) diffraction efficiency 
and spectral resolution. 

In conclusion, we have demonstrated the measurement of single-shot 
spectra of temporally selected micropulses from the MIR FEL by upconversion. 
With the microchip (Nd:YAG) laser we are able to obtain the single-shot 
spectra of only $\sim$2 (57) micropulses of MIR FEL at any temporal 
position within a macropulse. 
The spectral resolution of our current system is $\sim$7 cm$^{-1}$ 
($\sim$10 cm$^{-1}$) with the microchip (Nd:YAG) laser, and it is mostly
limited by the resolution of the spectrometer. 
If we use a spectrometer with a finer grating, say, 1200 lines/mm, 
we should be able to improve the resolution of the system to 
$\sim$1.2 cm$^{-1}$. 
The technique described in this work will be useful not only 
for the on-line monitoring of the FEL spectra but also for the 
investigation of the correlation between the properties of the electron 
beam and those of the FEL beam. 

\vspace*{2mm}
\noindent
We acknowledge Professor Tetsuo Sakka for the loan of spectrometer
and Tokyo Instruments Inc. for the loan of Nd:YAG laser. 
This work was supported by a Grant-in-Aid for scientific research from 
the Ministry of Education and Science of Japan.


\begin{thebibliography}{99}

\newcommand{\enquote}[1]{``#1''}

\bibitem{Deacon:1977zz}
D.~A. Deacon, L.~R. Elias, J.~M. Madey, G.~J. Ramian, H.~A. Schwettman, and
T.~I. Smith, Phys. Rev. Lett. \textbf{38}, 892 (1977).

\bibitem{OShea:2001gv}
P.~G. O'Shea and H.~P. Freund, Science \textbf{292}, 1853 (2001).

\bibitem{Zimdars:1993tf}
D.~Zimdars, A.~Tokmakoff, S.~Chen, S.~R. Greenfield, M.~D. Fayer, T.~I. Smith,
  and H.~A. Schwettman, Phys. Rev. Lett. \textbf{70}, 2718 (1993).

\bibitem{Shah:1988wl}
J.~Shah, IEEE J. Quantum Elect. \textbf{24}, 276 (1988).

\bibitem{Heilweil:1989tw}
E.~J. Heilweil, Opt. Lett. \textbf{14}, 551 (1989).

\bibitem{Kubarych:2005io}
K.~J. Kubarych, M.~Joffre, A.~Moore, N.~Belabas, and D.~M. Jonas, 
Opt. Lett. \textbf{30}, 1228 (2005).

\bibitem{DeCamp:2005va}
M.~DeCamp and A.~Tokmakoff, Opt. Lett. \textbf{30}, 1818 (2005).

\bibitem{Baiz:2011tf}
C.~Baiz and K.~Kubarych, Opt. Lett. \textbf{36}, 187 (2011).

\bibitem{Zhu:2012wq}
J.~Zhu, T.~Mathes, A.~D. Stahl, J.~T.~M. Kennis, and M.~L. Groot, 
Opt. Express  \textbf{20}, 10562 (2012).

\bibitem{Wijnen:2010tb}
F.~J.~P.~Wijnen, G.~Berden, and R.~T.~Jongma, Opt. Express \textbf{18}, 
26517 (2010).

\bibitem{kufel}
H. Ohgaki, T. Kii, K. Masuda, H. Zen, S. Sasaki,
T. Shiiyama, R. Kinjo, K. Yoshikawa, and T. Yamazaki,
Jpn. J. Appl. Phys. \textbf{47}, 8091 (2008). 

\end{thebibliography}
\end{document}